\documentclass{INTERSPEECH2023}

\interspeechcameraready

\usepackage{bbm}
\usepackage{graphicx}
\usepackage{amsmath}
\usepackage{amssymb}
\usepackage{booktabs}
\usepackage{enumitem}
\usepackage{rotating}
\usepackage{colortbl}
\usepackage[table]{xcolor}
\usepackage{makecell}
\usepackage{tabu}
\usepackage{subcaption}
\usepackage{multicol}
\usepackage{multirow}
\usepackage{csquotes}
\usepackage{wrapfig}

\title{Recycle-and-Distill: Universal Compression Strategy for Transformer-based Speech SSL Models with Attention Map Reusing and Masking Distillation}
\name{Kangwook Jang$^{1*}$\thanks{*\,These authors contributed equally to this work.}, Sungnyun Kim$^{2*}$, Se-Young Yun$^2$, Hoirin Kim$^1$}
\address{
  $^1$School of Electrical Engineering, KAIST \\
  $^2$Graduate School of AI, KAIST}
\email{\{dnrrkdwkd12, ksn4397, yunseyoung, hoirkim\}@kaist.ac.kr}

\begin{document}

\maketitle

\begin{abstract}
Transformer-based speech self-supervised learning (SSL) models, such as HuBERT, show surprising performance in various speech processing tasks.
However, huge number of parameters in speech SSL models necessitate the compression to a more compact model for wider usage in academia or small companies.
In this study, we suggest to reuse attention maps across the Transformer layers, so as to remove key and query parameters while retaining the number of layers.
Furthermore, we propose a novel masking distillation strategy to improve the student model's speech representation quality.
We extend the distillation loss to utilize both masked and unmasked speech frames to fully leverage the teacher model's high-quality representation.
Our universal compression strategy yields the student model that achieves phoneme error rate\,(PER) of 7.72\% and word error rate\,(WER) of 9.96\% on the SUPERB benchmark.
\end{abstract}

\noindent\textbf{Index Terms}: speech self-supervised learning, model compression, attention map reusing, masking distillation

\section{Introduction}
Transformer-based speech SSL models\,\cite{liu2020mockingjay, chi2021audio, liu2021tera} have been actively studied in speech processing field\,\cite{liu2022audio} as SSL arises as a successful representation learning approach in recent years\,\cite{mikolov2013efficient, caron2018deep, chen2020simple, baevski2022data2vec}.
Especially for wav2vec 2.0\,\cite{baevski2020wav2vec}, HuBERT\,\cite{hsu2021hubert}, and wavLM\,\cite{chen2021wavlm}, all of which are inherited from BERT\,\cite{devlin2018bert}, show surprising performance in automatic speech recognition\,(ASR), comparable to supervised learning approaches \cite{gulati2020conformer, kim2022squeezeformer}.
Since the versatility of speech SSL becomes also crucial, the above models have been further explored in various applications including automatic speaker verification (ASV)\,\cite{wang2021fine} or emotion recognition (ER)\,\cite{pepino2021emotion}. 

However, these models have huge number of parameters and are trained for very long time, which makes it hard for the resource-limited groups to train their own models.
For instance, wav2vec 2.0 \textsc{Large} with 317M parameters should be pretrained for more than 290 days on a single V100 GPU\,\cite{baevski2020wav2vec} on LibriSpeech dataset\,\cite{panayotov2015librispeech}.
This necessitates us to build a compressed model that allows much more parameter-efficient training and lower computational overhead. 

Knowledge distillation\,(KD)\,\cite{hinton2015distilling} is a common model compression technique where a smaller student model is being trained by distilling the knowledge from a teacher model.
Prior efforts in distilling large-scale speech SSL models have been made with reducing the number of Transformer layers or shrinking their width.
DistilHuBERT\,\cite{chang2021distilhubert} is distilled in a way of predicting multi-layer outputs of HuBERT, with most of the Transformer layers removed.
FitHuBERT\,\cite{lee2022fithubert}, instead of removing the layers, suggests cutting down the width of attention and feed-forward network\,(FFN) in each Transformer layer.
LightHuBERT\,\cite{wang2022lighthubert} creates a prunable supernet through distillation and conducts architecture search to make a small student.

Despite the effectiveness of previous approaches in mitigating the performance drop by compression, they still face several issues.
(1) Wide and shallow students\,\cite{chang2021distilhubert, ashihara2022deep} still exhibit degradation on content-related downstream tasks.
(2) Layer-to-layer\,(L2L) distillation is proved to be effective \cite{lee2022fithubert, ashihara2022deep}, however, it is counter-intuitive in terms of compression since every layer's parameters are required.
(3) Pruning by architecture search \cite{wang2022lighthubert} prepares an additional teacher-sized supernet using 32 GPUs, which is not end-to-end\,(E2E) and cannot be easily trained by resource-limited groups.

We suggest reusing attention maps across the student's Transformer layers, which is inspired by previous works\,\cite{xiao2019sharing, bhojanapalli2021leveraging} that claimed the similarity between attention maps.
Attention map reusing enables us to remove key and query parameters in certain Transformer layers, making it unnecessary to retain all layer parameters for L2L distillation.
Furthermore, we can reinvest the saved parameters to other parts of Transformer.

We also propose a masking with L2L distillation for better speech representation quality of our student model.
Masking speech frames is a widely used technique in speech SSL models\,\cite{baevski2020wav2vec, hsu2021hubert}, trained by predicting the masked representation.
This technique has been simply applied to distilling HuBERT\,\cite{wang2022lighthubert}, but not in the L2L manner.
Our novel masking distillation scheme aims to fully leverage the teacher's representation by extending the distillation loss to both masked and unmasked speech frames.
We emphasize that our scheme is an E2E fashion and enhances the general quality of speech representation, especially in content- and semantics-related tasks.

\begin{figure*}[!t]
    \centering
    \includegraphics[width=.72\linewidth]{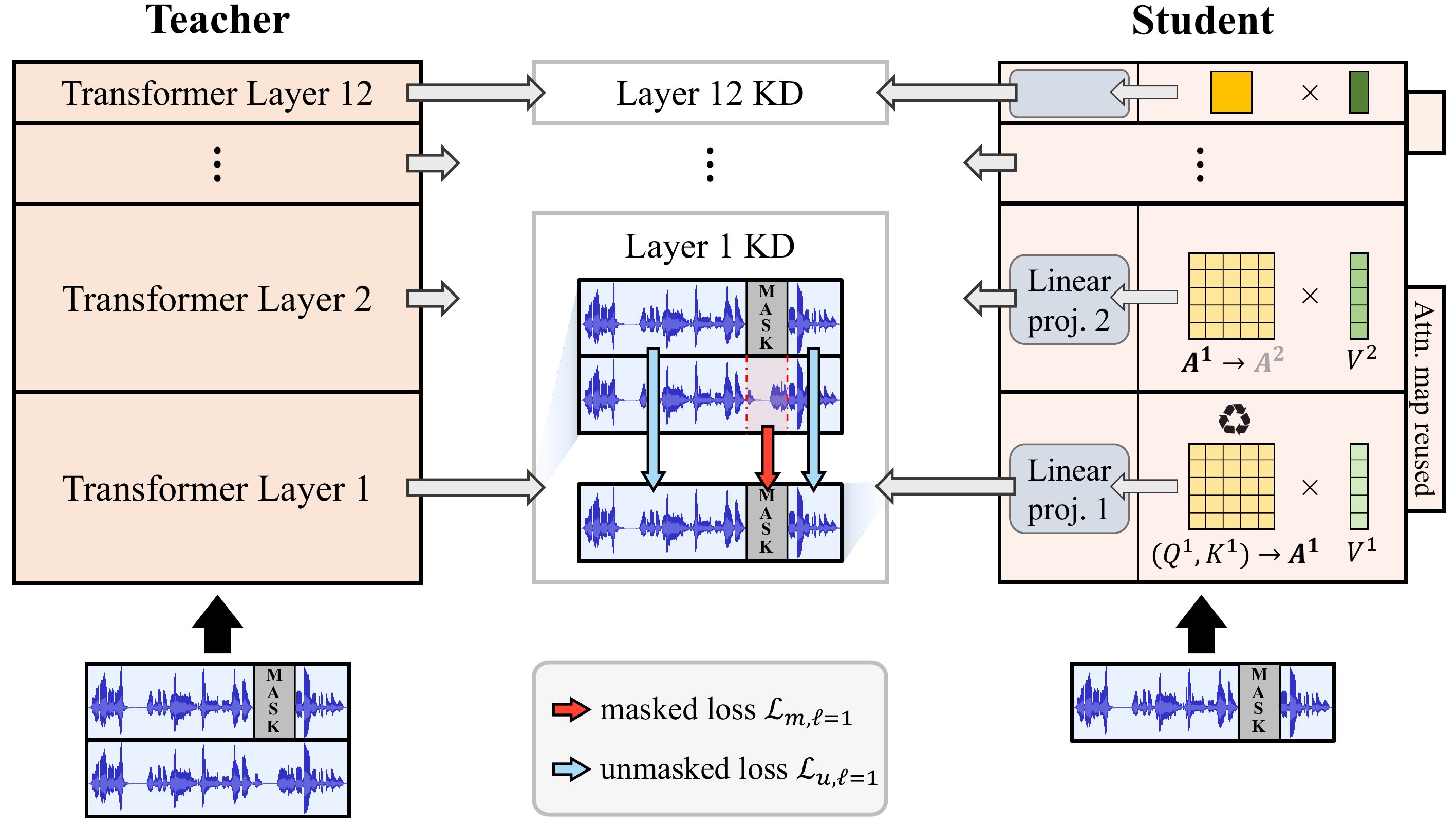}
    \caption{Our compression strategy involves reusing the attention map of the previous layer and extending the distillation process to masked\,(red arrow) and unmasked\,(blue arrow) representations.\,The input masked frames are identical for both teacher and student.}
    \label{fig:overview}
\vspace{-5pt}
\end{figure*}

Combining our two approaches described\,(Fig.\,\ref{fig:overview}), we reinvest the saved parameters from attention map reusing to FFN, and create our flagship model, \textbf{ARMHuBERT} (\underline{A}ttention map \underline{R}eused \underline{M}ask \underline{HuBERT}).
As evaluated on the SUPERB benchmark\,\cite{yang21c_interspeech}, 
ARMHuBERT achieves overall score\,\cite{chen2021wavlm} of 78.1, the state-of-the-art E2E distillation.
It also reaches 7.72\% PER in phoneme recognition\,(PR), and 9.96\% WER in ASR.

\section{Preliminaries}

\subsection{Transformer-based Speech SSL Models}
Recent dominant SSL models in speech field are wav2vec 2.0\,\cite{baevski2020wav2vec}, HuBERT \cite{hsu2021hubert}, and wavLM\,\cite{chen2021wavlm}, where these three model structures are identical except for detailed level.
Specifically, they share 12 or 24 Transformer\,\cite{vaswani2017attention} layers and 7-layer 1D-CNN.
Their pretraining schemes are based on masked prediction, estimating the codewords by output representation of the masked frames.
Despite the superiority and scalability of speech SSL models, large number of parameters and their computational overhead make it difficult to train these models.
We thus implement the model compression on HuBERT and wavLM, the two dominant SSL models in speech, to demonstrate the effectiveness of our compression strategy.

\subsection{SUPERB Benchmark}
The beginning of speech SSL models focused on content-related downstream tasks such as ASR or PR \cite{oord2018representation, schneider2019wav2vec}, however, their versatility to other tasks has been recognized as crucial recently\,\cite{chen2021wavlm}.
In this context, SUPERB benchmark\,\cite{yang21c_interspeech} has been proposed to evaluate the generalizability of speech SSL models, covering the aspects of content, speaker, semantics, and paralinguistics.
We evaluate our representation against the SUPERB benchmark to verify the generalizability of our student model.
The SUPERB downstream tasks include PR, ASR, keyword spotting\,(KS), query-by-example spoken term detection\,(QbE), speaker identification\,(SID), ASV, speaker diarization\,(SD), intent classification\,(IC), slot filling\,(SF), and ER.

\section{Methodology}

\subsection{Attention Map Reusing}

\begin{table*}[!t]
    \centering
    \caption{Evaluation results on SUPERB benchmark. Metrics include parameter size in million, PER\%, WER\%\,(w/o language model), accuracy\,(Acc\%), maximum term weighted value (MTWV), equal error rate (EER\%), diarization error rate (DER\%), F1 score (F1\%), and concept error rate (CER\%). ``Overall'' denotes the average scoring of all tasks proposed in \cite{chen2021wavlm}. LightHuBERT\,\cite{wang2022lighthubert} works by two-stage training, where HuBERT-sized supernet needs to be trained first, thus not compared with E2E distillation models. ARMHuBERT-S and ARMwavLM-S with 960h distillation are trained for 100 epochs.}
    \label{tab:superb_benchmark}
    \vspace{-5pt}
    \addtolength{\tabcolsep}{-1pt}
    \resizebox{\textwidth}{!}{
    \begin{tabular}{lccccccccccccc}
    \toprule
    & & & \multicolumn{4}{c}{Content} &  \multicolumn{3}{c}{Speaker} & \multicolumn{3}{c}{Semantics} & Paral. \\
    \cmidrule(l{2pt}r{2pt}){4-7} \cmidrule(l{2pt}r{2pt}){8-10} \cmidrule(l{2pt}r{2pt}){11-13} \cmidrule(l{2pt}r{2pt}){14-14}
    & Params & Overall & PR & ASR & KS & QbE & SID & ASV & SD & IC & \multicolumn{2}{c}{SF} & ER \\
    \cmidrule(l{2pt}r{2pt}){2-2} \cmidrule(l{2pt}r{2pt}){3-3} \cmidrule(l{2pt}r{2pt}){4-4} \cmidrule(l{2pt}r{2pt}){5-5} \cmidrule(l{2pt}r{2pt}){6-6} \cmidrule(l{2pt}r{2pt}){7-7} \cmidrule(l{2pt}r{2pt}){8-8} \cmidrule(l{2pt}r{2pt}){9-9} \cmidrule(l{2pt}r{2pt}){10-10} \cmidrule(l{2pt}r{2pt}){11-11} \cmidrule(l{2pt}r{2pt}){12-13} \cmidrule(l{2pt}r{2pt}){14-14}
    Models & \negthickspace Millions\,$\downarrow$\negthickspace  & Score\,$\uparrow$ & PER\,$\downarrow$ & WER\,$\downarrow$ & Acc\,$\uparrow$ & MTWV\,$\uparrow$ & Acc\,$\uparrow$ & EER\,$\downarrow$ & DER\,$\downarrow$ & Acc\,$\uparrow$ & F1\,$\uparrow$ & CER\,$\downarrow$ & Acc\,$\uparrow$ \\
    \midrule
    \textit{\textcolor{gray}{Baselines}}  & \\
    FBANK\,\cite{yang21c_interspeech} & 0 & 40.5 & 82.01 & 23.18 & 8.63 & 0.0058 & 8.5E-4 & 9.56 & 10.55 & 9.1 & 69.64 & 52.94 & 35.39 \\
    HuBERT\,\textsc{Base} \cite{hsu2021hubert} & 94.70 & 80.8 & 5.41 & 6.42 & 96.30 & 0.0736 & 81.42 & 5.11 & 5.88 & 98.34 & 88.53 & 25.20 & 64.92 \\
    wavLM\,\textsc{Base} \cite{chen2021wavlm} & 94.70 & 81.9 & 4.84 & 6.21 & 96.79 & 0.0870 & 84.51 & 4.69 & 4.55 & 98.63 & 89.38 & 22.86 & 65.94 \\
    LightHuBERT $a_\text{Small}$\,\cite{wang2022lighthubert} & \negthickspace 94.7\,$\rightarrow$\,27.00 \negthickspace & 79.1 & 6.60 & 8.33 & 96.07 & 0.0764 & 69.70 & 5.42 & 5.85 & 98.23 & 87.58 & 26.90 & 64.12 \\
    \midrule
    \multicolumn{5}{l}{\textit{\textcolor{gray}{960h distillation -- \# params: 26.4M\,$\sim$\,31.6M}}} & \\
    FitW2V2\,\cite{lee2022fithubert} & 31.63 & 76.5 & 12.22 & 11.44 & 96.04 & 0.0475 & 64.71 & 6.65 & 6.44 & 93.38 & 86.65 & 29.40 & 62.35 \\
    3-L \textsc{One}-Pred\,\cite{ashihara2022deep} & 30.58 & 76.8 & 13.34 & 12.23 & 96.69 & 0.0489 & 75.71 & 6.48 & 6.56 & 94.15 & 82.89 & 34.65 & 63.95 \\
    12-L \textsc{Half}-L2L\,\cite{ashihara2022deep} & 26.87 & 77.6 & 10.67 & 10.96 & 97.24 & 0.0604 & 69.52 & 6.13 & 6.81 & 96.97 & 86.11 & 30.93 & 63.24 \\
    \textbf{MaskHuBERT (ours)} & 26.64 & 77.8 & \textbf{7.30} & \textbf{9.77} & 96.36 & \textbf{0.0664} & 62.83 & \textbf{5.38} & 6.79 & 97.05 & 87.31 & 27.10 & 62.37 \\
    \textbf{ARMHuBERT (ours)} & \textbf{26.45} & \textbf{78.1} & 7.72 & 9.96 & 96.88 & 0.0635 & 65.03 & 5.68 & 7.10 & \textbf{97.07} & \textbf{87.59} & \textbf{26.06} & 62.86 \\
    \midrule
    \multicolumn{5}{l}{\textit{\textcolor{gray}{960h distillation -- \# params: 22.4M\,$\sim$\,23.5M}}} & \\
    DistilHuBERT\,\cite{chang2021distilhubert} & 23.49 & 75.9 & 16.27 & 13.37 & 95.98 & 0.0511 & 73.54 & 8.55 & 6.19 & 94.99 & 82.57 & 35.59 & 63.02 \\
    FitHuBERT\,\cite{lee2022fithubert} & 22.49 & 74.5 & 13.32 & 12.09 & 96.27 & 0.0489 & 55.71 & 8.00 & 6.84 & 91.25 & 84.06 & 32.46 & 59.82 \\
    \textbf{ARMHuBERT-S (ours)} & \textbf{22.39} & 77.5 & 8.63 & 10.82 & 96.82 & 0.0720 & 63.76 & \textbf{5.58} & 7.01 & 97.02 & 86.34 & 29.02 & 62.96 \\
    \textbf{ARMwavLM-S (ours)} & \textbf{22.39} & \textbf{78.9} & \textbf{7.42} & \textbf{10.03} & \textbf{97.01} & \textbf{0.0741} & 71.29 & 5.99 & 7.11 & \textbf{97.76} & \textbf{87.41} & \textbf{26.97} & \textbf{64.54} \\
    \midrule
    \multicolumn{5}{l}{\textit{\textcolor{gray}{100h distillation}}} & \\
    FitW2V2\,\cite{lee2022fithubert} & 22.49 & 73.1 & 16.50 & 14.77 & 94.68 & 0.0380 & 51.65 & 7.43 & 6.94 & 90.03 & 81.95 & 34.74 & 62.87 \\
    FitHuBERT\,\cite{lee2022fithubert} & 22.49 & 74.5 & 14.05 & 12.66 & 96.23 & 0.0579 & 54.24 & 7.88 & 7.19 & 94.20 & 83.41 & 34.00 & 61.67 \\
    \textbf{ARMHuBERT-S (ours)} & \textbf{22.39} & 76.8 & 9.17 & 11.83 & 96.01 & 0.0569 & \textbf{66.48} & \textbf{5.92} & \textbf{6.23} & 95.97 & 83.89 & 33.29 & 63.29 \\
    \textbf{ARMwavLM-S (ours)} & \textbf{22.39} & \textbf{77.0} & \textbf{8.33} & \textbf{11.37} & \textbf{96.30} & \textbf{0.0579} & 65.40 & 6.38 & 7.41 & \textbf{96.76} & \textbf{84.89} & \textbf{31.95} & \textbf{63.41} \\
    \bottomrule
    \end{tabular}
    }
\vspace{-5pt}
\end{table*}
Attention map reusing is a technique for substituting the present layer's attention map with the previous one, which has been covered in several domains\,\cite{xiao2019sharing, shim2022understanding}.
Prior works\,\cite{xiao2019sharing, bhojanapalli2021leveraging} have pointed out the similarity of the attention maps across heads and layers in pretrained Transformer models, such as BERT\,\cite{devlin2018bert} and ViT\,\cite{dosovitskiy2020image}.
We leverage this property by reusing the attention maps to compress the student model.
Alternatively, we can reassign the amount of parameters saved by attention map reusing, without increasing the total number of parameters.

In Transformer's multi-head self-attention\,(MHSA) module\,\cite{vaswani2017attention}, the input $x \in \mathbb{R}^{n \times d}$ with the sequence length $n$ is transformed to $H$ independent queries, keys, and values by transformation matrices $W_{h, k}, W_{h, q} \in \mathbb{R}^{d \times d_{k}}$, and $W_{h, v} \in \mathbb{R}^{d \times d_{v}}$, respectively, for each head $h$.
Here, $d_{k}$, $d_{v}$, and $d$ are the width of the keys, values, and model, respectively.

\begin{equation}\label{eq:self-attn}
\begin{array}{cc}
    K_h = W_{h, k}x, & K_h \in \mathbb{R}^{n \times d_k}, \\
    Q_h = W_{h, q}x, & Q_h \in \mathbb{R}^{n \times d_k}, \\
    V_h = W_{h, v}x, & V_h \in \mathbb{R}^{n \times d_v}
\end{array}
\end{equation}
Then, key and query are multiplied along the width axis to obtain a scaled dot-product attention map, $A_h \in \mathbb{R}^{n \times n}$.
Linear combinations of the attention map and value for each head are concatenated, and then projected to the original width.
\begin{gather}\label{eq:attn_map}
    A_h = \textit{softmax}\big( Q_h K^{\top}_h / \sqrt{d_k} \big), \\ 
    \textit{MHSA}(x) = \big[A_{1}V_{1}, \dots, A_{H}V_{H} \big]W_o,\,\, W_o \in \mathbb{R}^{Hd_v \times d}
\end{gather}

Attention map reusing is to replace $A_h$ with the previous layer's one.
For instance, if we reuse the $k$-th previous attention map on the current layer $\ell$, the ReuseMHSA module is
\begin{equation}\label{eq:reuse_attn}
\begin{array}{c}
    \textit{ReuseMHSA}(x) = \big[{A_1^{\ell-k}}{V_1^{\ell}}, \dots ,{A_H^{\ell-k}}{V_H^{\ell}}\big]\,{W_o^{\ell}}. \\
\end{array}
\end{equation}

\noindent Accordingly, computing $K_h$ and $Q_h$ can be omitted, reducing the number of multiplications and additions by $(2nd^2+n^2d)$.
Assuming $d/H$\,$=$\,$d_v$\,$=$\,$d_k$, the omitted computation accounts for half of the original computation for MHSA, which is $(4nd^2+2n^2d)$.
As a result, less parameters and multiply-accumulates\,(MACs) are required as more ReuseMHSA modules are employed (see Sec.\,\ref{subsec:where_to_reuse}). 

\subsection{Masking Distillation}
\label{sec:masking_distillation}
Attention map reusing has reduced the number of parameters, however, it may affect the representation quality of the student model.
To improve the student's representation learning, we offer a novel masking distillation scheme that leverages the teacher's representation knowledge in a more sophisticated way.

Speech frame masking involves learning representation through masked prediction, where the model learns to represent masked frames accurately based on other unmasked frames.
LightHuBERT\,\cite{wang2022lighthubert}, inspired by data2vec\,\cite{baevski2022data2vec}, has first applied the masking strategy to distilling HuBERT.
In this approach, the teacher model guides the representation of masked frames.
Let $\mu(x)$ be the masked input, and $f^t$ and $f^s$ the teacher and student model.
Then, the masked loss function becomes
\begin{equation}\label{eq:lighthubert}
    \mathcal{L}(x) = \frac{1}{|M|}\sum_{i \in M} \big\lVert f^t_i(x) - f^s_i(\mu(x)) \big\rVert_2
\end{equation}
where $f_i$ is the $i$-th frame of the speech representation, and $M$ is the set of the masked frames.

In addition to the masked part loss\,(eq.\,\ref{eq:lighthubert}), we suggest to employ an unmasked loss since the teacher model can provide high-quality representation even on the unmasked frames.
However, if the masking process removes essential frames, distilling the intact form of $f^t(x)$ can leak such essential knowledge that should have been removed.
This induces biased predictions of the student, as it learns information that cannot be inferred from the masked input.

To prevent this, we make the teacher model receive the same masked input as the student does when distilling the unmasked part. Hence, the entire distillation loss becomes 
\begin{align}\label{eq:mask_distill}
    \mathcal{L}(x) &=  \sum_{\ell} \alpha_\ell \big[ \mathcal{L}_{m,\ell}(x) + \mathcal{L}_{u,\ell}(x) \big] 
    \nonumber \\
    &= \sum_{\ell} \frac{\alpha_\ell}{|M|} \sum_{i \in M} \big\lVert f^t_{i,\ell}(x) - f^s_{i,\ell}(\mu(x)) \big\rVert_2 \\
    &+ \sum_{\ell}\frac{\alpha_\ell}{n-|M|}\sum_{i \notin M} \big\lVert f^t_{i,\ell}(\mu(x)) - f^s_{i,\ell}(\mu(x)) \big\rVert_2 \nonumber 
\end{align}
where $\alpha_\ell$ is the layerwise coefficient.
$\mathcal{L}_{m,\ell}$ and $\mathcal{L}_{u,\ell}$ represent masked loss and unmasked loss of the $\ell$-th layer, respectively.

In summary, our novel masking distillation strategy appropriately guides the student's knowledge acquisition, by distilling not only the masked representation of unmasked data but also the unmasked representation of masked data (see Fig.\,\ref{fig:overview}).
In Sec.\,\ref{subsec:how_to_mask}, we investigate the strength of our masking strategy compared to other types of losses.

\section{Results}

\subsection{Implementation Details}
We distilled the two dominant Transformer-based speech SSL models, HuBERT \textsc{Base}\,\cite{hsu2021hubert} and wavLM \textsc{Base}\,\cite{chen2021wavlm}, that are pretrained on LibriSpeech 960 hours dataset\,\cite{panayotov2015librispeech}.
Our student model consists of 12 layers of Transformers as the teachers, while the detailed design mostly follows FitHuBERT\,\cite{lee2022fithubert}: width of attention and FFN reduced and linear projections adopted at each layer.
The layerwise coefficients $\alpha_\ell$ are set to 0.1 except for the last layer, where it is set to 1.
Unless specified, the LibriSpeech\,\cite{panayotov2015librispeech} dataset is distilled for 200 epochs with effective batch size of 72 including gradient accumulation.

\vspace*{3pt}
\noindent\textbf{Reuse pattern}\quad We employ an alternating reuse pattern for the attention maps, whereby the attention map of an even-numbered Transformer layer is repeated by that of the previous odd-numbered layer.
We denote this pattern as \textit{2by6}, our default setting. We examine other reuse patterns in Sec.\,\ref{subsec:where_to_reuse} in terms of performance, number of parameters, and MACs.

\vspace*{3pt}
\noindent \textbf{Model description}\quad To verify our masking distillation strategy, we first build a student model, MaskHuBERT, which employs masking distillation only.
MaskHuBERT has the width of (attention, FFN) as (480, 640).
Then, \textit{2by6} reuse pattern is applied to MaskHuBERT, leading to 10.3\% of parameter reduction.
We extend this model to two options: ARMHuBERT and ARMHuBERT-S.
ARMHuBERT is a reinvested version of MaskHuBERT, where the saved parameters from attention map reusing are reassigned to FFN, resulting in increased width of (480, 864).
ARMHuBERT-S is a reduced version to match the parameters with previous works, having the width of (432, 816).
To establish the universality of our strategy, we introduce ARMwavLM-S that is structurally identical to ARMHuBERT-S, with the only change in teacher from HuBERT to wavLM.

\subsection{SUPERB Benchmark Results}
\label{subsec:superb_results}
In Table\,\ref{tab:superb_benchmark}, we evaluate our student models on SUPERB benchmark\,\cite{yang21c_interspeech}.
We follow the default fine-tuning recipes, including a learning rate scheduler, with the learning rate scaled to 10$\times$ in SID task.
MaskHuBERT outperforms 12-L \textsc{Half}-L2L, the previous state-of-the-art E2E distillation method, with less parameters used.
Our observation indicates that incorporating our masking strategy into the L2L distillation\,\cite{lee2022fithubert, ashihara2022deep} results in enhancing the student's representation quality.
Especially, MaskHuBERT highly improves the performances in content- and semantics-related tasks.

ARMHuBERT achieves a better overall score of 78.1 with less parameters than MaskHuBERT.
Despite the removal of certain attention parameters, increasing the FFN width contributes to better quality of speech representation, achieving 7.72\% PER and 9.96\% WER. 
We find out that ARMHuBERT shows promising improvements when compared to MaskHuBERT in SF and SID tasks, exhibiting a similar level of performance in other tasks.
In the end, the number of parameters and MACs in ARMHuBERT have decreased to 28\% and 30\% of the teacher model, HuBERT \textsc{Base}\,\cite{hsu2021hubert}, respectively.

In a smaller parameter group, ARMHuBERT-S, the param-eter-reduced version, outperforms DistilHuBERT and FitHuBERT by a large margin.
Specifically, ARMHuBERT-S also shows the outstanding results in content- and semantics-related tasks,
which means the consistency of the representations produced by MaskHuBERT and ARMHuBERT-S.
In addition, the result that ARMwavLM-S surpasses ARMHuBERT-S implies the universality of our strategy: without any modifications of the student model structure, replacing with a superior teacher model creates a better student.
The results of the LibriSpeech\,\cite{panayotov2015librispeech} 100h distillation are also consistent with the formerly demonstrated results.

\section{Discussions}

\begin{table}[!t]
    \centering
    \small
    \caption{Performance comparisons of various reusing patterns. Parameter size\,(M) and MACs\,(G) are additionally measured. The width of (attention, FFN) for each model is (432, 816), while ``-up'' suffix denotes more parameters assigned to FFN to match with \textit{2by6}.
    Masking is not applied here.}
    \label{tab:attention_reusing}
    \vspace{-5pt}
    \addtolength{\tabcolsep}{-3pt}
    \resizebox{\linewidth}{!}{
    \begin{tabular}{llcccccc}
    \toprule
    pattern & reused layers & \negthickspace params \negthickspace & MACs & WER\,$\downarrow$ & EER\,$\downarrow$ & F1\,$\uparrow$ & CER\,$\downarrow$ \\
    \midrule
    \textit{6by2} & \{1,7\} & 20.90 & 423 & 13.52 & 6.30 & 83.69 & 34.92 \\
    \textit{3by4}& \{1,4,7,10\} & 21.65 & 437 & 12.37 & \textbf{5.67} & 83.29 & 33.60 \\
    \textit{6by2-up}& \{1,7\} & 22.39 & 440 & 13.18 & 5.89 & 83.07 & 34.79 \\
    \textit{3by4-up}& \{1,4,7,10\} & 22.39 & 445 & 12.39 & 6.06 & 83.79 & 33.49 \\
    \textit{2by6} & \{1,3,5,7,9,11\}\! & 22.39 & 450 & \textbf{12.18} & 5.95 & \textbf{84.91} & \textbf{32.29} \\
    \midrule
    None & - & 24.64 & 490 & 11.94 & 5.87 & 84.78 & 31.38 \\
    \bottomrule
    \end{tabular}
    }
    \vspace{-5pt}
\end{table}

In this section, we explore which layer's attention map should be reused in other layers and how to implement the masking distillation.
Unless specified, we have conducted the distillations on 100-hour of LibriSpeech\,\cite{panayotov2015librispeech} and evaluated on the ASR, ASV, and SF tasks of the SUPERB benchmark\,\cite{yang21c_interspeech}.

\subsection{Where to Reuse}
\label{subsec:where_to_reuse}
Table\,\ref{tab:attention_reusing} summarizes the performance depending on various attention map reusing patterns, and in general, the \textit{2by6} pattern performs the best.
Other reuse patterns have reduced the Transformer's representation capacity due to overly frequent reusing.
Assigning more parameters to FFN (\textit{-up}) still has limit in terms of the performance gain.
Comparing to no reuse pattern applied, the performance decrease of \textit{2by6} is small, but it takes advantages in 9.13\% and 8.16\% reduction of parameters and MACs, respectively.
We note that the number of MACs in a single reuse MHSA module\,(eq.\,\ref{eq:reuse_attn}) is reduced by half, from 13.2G to 6.6G.

\subsection{How to Mask}
\label{subsec:how_to_mask}

\noindent\textbf{Masking strategy}\quad 
Table\,\ref{tab:masking_ablation} shows the efficacy of our masking strategy.
We first eliminated the loss function on the unmasked frames\,($\mathcal{L}_{u,\ell}$), making it equivalent to the L2L version of the LightHuBERT\,\cite{wang2022lighthubert} distillation loss. This approach severely damaged performances, particularly in the ASR and ASV tasks. Next, we modified the unmasked loss function to distill from the unmasked input, \textit{i.e.},\,only $f^t(x)$ being distilled to the student. This also led to degraded performance in most tasks, revealing that our unmasked loss with masked input properly guides the knowledge acquisition without imposing biased predictions.

\vspace*{3pt}
\noindent\textbf{Masking ratio}\quad 
High value of masking ratio can lead to a student model producing good representation, as it has less information to infer with\,\cite{hsu2021hubert, he2022masked}.
However, it can also make the learning process more difficult.
In Table\,\ref{tab:masking_ratio}, we examine the optimal masking ratios for each training set.
For LibriSpeech 960h\,\cite{panayotov2015librispeech}, both ratios of 0.4 and 0.8 produce excellent results.
On the other hand, for the 100h dataset, ratio of 0.4 produces the best results overall.
This implies that lower masking ratio is preferred in low-resource distillation setting.
Accordingly, in our main experiments, we have used the ratios of 0.8 and 0.4 for the 960h and 100h distillation, respectively.

\begin{table}[!t]
    \centering
    \small
    \caption{Ablation study on our masking strategy.}
    \label{tab:masking_ablation}
    \vspace{-5pt}
    \resizebox{\linewidth}{!}{
    \begin{tabular}{lcccc}
    \toprule
    methods & WER\,$\downarrow$ & EER\,$\downarrow$ & F1\,$\uparrow$ & CER\,$\downarrow$ \\
    \midrule
    MaskHuBERT-100h & \textbf{11.56} & \textbf{5.87} & \textbf{84.31} & 32.28 \\
    \,[--] distil. unmasked part & 13.23 & 7.96 & 82.78 & 33.53 \\
    \,[--] distil. from masked input & 11.65 & 6.09 & 84.29 & \textbf{31.41} \\
    \bottomrule
    \end{tabular}
    }
\end{table}

\begin{table}[!t]
    \centering
    \small
    \caption{Performance comparisons with different masking ratios.\,``sch''\,indicates linear scheduling of the ratio as 0.4 to 0.8.} 
    \label{tab:masking_ratio}
    \vspace{-5pt}
    \resizebox{\linewidth}{!}{
    \begin{tabular}{lccccc}
    \toprule
    models & ratio & WER\,$\downarrow$ & EER\,$\downarrow$ & F1\,$\uparrow$ & CER\,$\downarrow$ \\
    \midrule
    MaskHuBERT-960h & 0.4 & \textbf{9.75} & 5.58 & 86.94 & \textbf{26.79} \\
    MaskHuBERT-960h & 0.8 & 9.77 & \textbf{5.38} & \textbf{87.31} & 27.10 \\
    \midrule
    MaskHuBERT-100h & 0.4 & \textbf{11.56} & \textbf{5.87} & \textbf{84.31} & \textbf{32.28} \\
    MaskHuBERT-100h & 0.6 & 11.99 & 6.18 & 83.42 & 33.31 \\
    MaskHuBERT-100h & 0.8 & 12.74 & 6.56 & 83.68 & 33.82 \\
    MaskHuBERT-100h & sch & 12.07 & 6.29 & 83.84 & 33.50 \\
    \bottomrule
    \end{tabular}
    }
    \vspace{-5pt}
\end{table}

\section{Conclusion and Future Work}
In summary, we have proposed the universal compression strategy which involves attention map reusing and novel masking distillation.
Our parameter-reinvested model, ARMHuBERT, achieves great performance in content- and semantics-related tasks.
Our strategy can be applied to any Transformer-based speech SSL models, and contributes to enhancing the general quality of speech representation.
Future work can focus on further improving our model on speaker-related tasks.

\section{Acknowledgements}
The study was supported by Korea Health Technology R\&D Project through the Korea Health Industry Development Institute funded by the Ministry of Health and Welfare, Republic of Korea (HR18C0016).

\clearpage
\newpage

\bibliographystyle{IEEEtran}
\bibliography{mybib}

\end{document}